\documentclass[11pt]{article}
\usepackage{graphicx}
\usepackage{setspace}
\usepackage{array}

\usepackage{amsmath}
\usepackage{authblk}
\usepackage{epstopdf,float}
\usepackage{authblk,multirow}
\usepackage{cite}

\usepackage{fullpage}
\title{\bf An Analytical Study on the Quasiperiodic Route to Chaos in a Forced Negative Conductance Circuit}
\author{G. Sivaganesh \footnote{sivaganesh.nld@gmail.com}}
\affil{Department of Physics, Alagappa Chettiar College of Engineering $\&$ Technology, Karaikudi-630 004, India}
\date{\today}
\begin{document}
\maketitle
\begin{center}
\bf Abstract
\end{center}
The quasiperiodic route to chaos in a piecewise linear forced parallel LCR circuit with a negative conductance and diode is studied analytically. An explicit analytical solution for the normalized state equations of the piecewise linear circuit is presented explaining the quasiperiodic route to chaos through phase portraits and power spectrum.
\begin{flushleft}
$\bf{Keywords:}$ quasiperiodic, chaos \\
$\bf{PACS:}$ 05.45.-a
\end{flushleft}

\section{Introduction}
Ever since the discovery of the Chua's circuit \cite{lab1}, chaotic dynamics has been observed in a variety of nonlinear electronic circuits. The chaotic attractor exhibited by the Chua's circuit was studied analyticaly and numerically \cite{lab2}.The circuit dynamics thus observed through experimental studies are well supported by numerical studies. Several routes to chaos such as the period doubling route to chaos or the Feignbaum scenario, torus breakdown to chaos or quasiperiodic route to chaos etc., are observed\cite{lab3,lab4}. The chaotic nature of the attractors thus obtained is confirmed by the presence of a positive lyapunov exponent. Chaotic circuits find potential applications in the field of secure communication through the concept of chaos synchronization. \\
Over the past three decades several piecewise linear autonomous and nonautonomous dissipative electronic circuits have been identified to possess chaotic attractors. Higher order systems $(order \ge 3)$ are mostly studied experimentally and numerically while some are studied analytically. However, the explicit analytical solutions thus obtained are not used to study the chaotic dynamics of the system\cite{lab5,lab6,lab7}.\\   
The circuit equations of second order non-autonomous chaotic circuits have mathematical simplicity to be solved analytically. The chaotic dynamics of some of the second order systems are studied through phase portraits obtained from the explicit analytical solutions of the normalized circuit equations\cite{lab8,lab9,lab10,lab11,lab12,lab13,lab14}. These second order systems have piecewise linear circuit circuit elements as the only nonlinear element. Hence the normalized circuit equations can be solved for each of the piecewise linear regions. Analytical study through phase portraits is carried out in systems exhibiting period doubling route to chaos and quasiperiodic route to chaos.\\ 
Chaos through torus breakdown in a piecewise linear forced parallel $LCR$ circuit with a diode and negative conductance was studied numerically by Inaba and Mori\cite{lab4}. An explicit analytical solution was presented to the normalized circuit equations in their study. However, the circuit dynamic was not studied using the solution presented. The forced series $LCR$ circuit with a diode and negative conductance connected parallel to the capacitor was introduced by Thamilmaran \cite{lab10}. The circuit exhibits both torus break down to chaos of {\emph{inaba-mori}} type \cite{lab4} and period doubling route to chaos of {\emph{Murali-Lakshmanan-Chua(MLC)}} types \cite{lab8}. An explicit analytical solution to the normalized circuit equations was presented and are used to draw phase portraits of chaotic attractors \cite{lab10}. \\
In the present work we give an explicit analytical solution to the normalized circuit equations of the forced parallel $LCR$ circuit with a diode and negative conductance that exhibits torus breakbown or qusiperiodic route to chaos. The resulting solution thus obtained is plotted to get phase portraits and power spectrum which reveals thequasiperiodic route to chaos.

\section{Circuit Equations}

\begin{figure}[H]
\begin{center}
\includegraphics[scale=0.5]{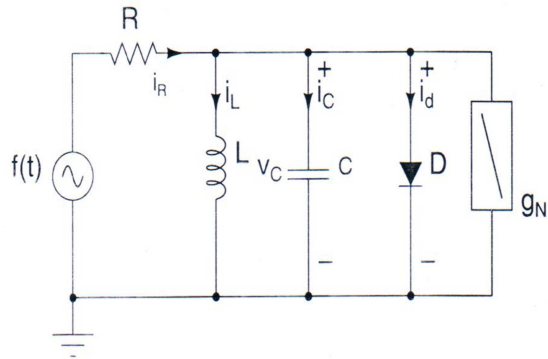}
\caption{Schematic diagram of the forced parallel LCR circuit with a diode and negative conductance }
\label{fig1}
\end{center}
\end{figure}

The schematic diagram of the piecewise linear forced parallel $LCR$ circuit with a negative conductance and diode is as shown in Fig.\ref{fig1}. The circuit consists of a simple parallel {\emph{LCR}} network, forced by a sinusoidal voltage generators $ F sin(\omega t)$  with a diode and negative conductance connected parallel to the Capacitor $C$. By applying Kirchhoff's voltage and current laws the voltage $v$ across the capacitor $C$ and the current $i_L$ through the inductor $L$ are  are given by the following set of two first-order coupled nonautonomous differential equations.

\begin{subequations}
\begin{eqnarray} 
L {di_L \over d\tau } & = &  v ,\\
C {dv \over d\tau } & = & -  i_L + (g_N - R)v -  i_D + R~F sin( \Omega \tau)
\vspace{-1.5cm}
\end{eqnarray}
\end{subequations}
where 
\begin{equation}
i_{D}(v) =
\begin{cases}
g_{D}~(v-V) & \text{if $v\ge V$}\\
0 & \text{if $v < V$}\\
\end{cases}
\end{equation}
where $g_{D}$ is slope of the characteristic curve of the diode. $F$ and $\Omega$ are the amplitude and frequency of the external sinusoidal force. The circuit exhibits torus breakdown route or quasiperiodic route to chaos for the circuit parameters taking the values  $C=8.067 nF, L=499 mH, g_D=12820 \mu S, g_N=139.8 \mu S, R=29516 \Omega$, break voltage $V=0.5 V$, $\Omega=752 Hz$.
\vspace{-0.20cm}

\section{\bf Explicit analytical solutions}
\vspace{-0.10cm} 
In this section, we will investigate the analytical aspects of the forced parallel $LCR$ circuit shown in Fig\ref{fig1}. Eqs.(1) can be written in the dimensionless form as, 
\begin{subequations}
\begin{eqnarray}
\dot x & = & y , \\
\dot y & = & - f(y) - x + f sin( \omega t), 
\end{eqnarray}
\end{subequations}
where
\begin{equation}
f(y) =
\begin{cases}
\alpha (y-1) - 2 \delta, & \text{if $y\ge 1$}\\
-2 \delta y,  & \text{if $y < 1$}\\
\end{cases}
\end{equation}
where, $t=(\tau/ \sqrt{LC}), x=\frac{i_L}{V} \sqrt(L/C), y=\frac{v}{V} \sqrt{LC}, \delta= \frac{g_N - R}{2} \sqrt(L/C), \alpha = (g_D+R-g_N) \sqrt(L/C), f=\frac{RF}{V} \sqrt(L/C), \omega_1= \Omega \sqrt{LC}$. The values of the rescaled parameters are $\delta=0.4, \alpha=100, \omega=0.3$. The amplitude of the external force $f$ is taken as the control parameter.
The analytical solution can be found by solving the set of normalized equations given in Eqs.~(3) for each of the piecewise linear regions given in Eq.(4), for the chosen circuit parameter values used in the experiment.

One can easily establish that a unique equilibrium point $ (x_0,y_0)$ exists for equation $(3)$ in each of the following two subsets,\\
\begin{equation}
\left.
\begin{aligned}
D_{On} & =  \{ (x^{*},y^{*})| x \ge 1 \} D^+ = (\alpha+2 \delta,0)\\
D_{Off} & =  \{ (x^{*},y^{*})|| x | < 1 \}| D^- = (0,0)
\end{aligned}
\right\}
\quad\text{}
\end{equation}
Hence the eqlibrium point in the $D_{On}$ region is $(\alpha+2 \delta,0)$ while in the $D_{Off}$ region it is the origin $(0,0)$. The stability of the equilibrium points given in equation $(5)$, can be calculated from the stability matrices.\\ 

$Region$ $D_{On}$ 
\\

In this region the stability determining eigen values are calculated from the stability matrix
\\
\begin{equation}
J_{On} =
\begin{pmatrix}
0 &&& 1 \\
-1 &&& -\alpha \\
\end{pmatrix}
\end{equation}
and the eigen values are real and distinct given by $m_{1,2}=-0.010001, -99.989998$. Hence the equilibrium point $(\alpha+2 \delta,0)$ corresponding to the $D_{On}$ region is a {\emph{stable node equilibrium point}}.
\\

$Region$ $D_{Off}$ 
\\

The stability determining eigen values for this region are calculated from the stability matrix
\\
\begin{equation}
J_{Off} =
\begin{pmatrix}
0 &&& 1 \\
-1 &&& 2 \delta \\
\end{pmatrix}
\end{equation}
and the eigen value are a pair of complex conjugates with a positive real part given by $m_{3,4}=0.4 \pm i 0.916515$. Hence the equilibrium point $(0,0)$ corresponding to the $D_{Off}$ region is an {\emph{unstable focus equilibrium point}}.\\
Naturally, these fixed points can be observed depending upon the initial conditions $x(0)$ and $y(0)$ of Eqs.~(3) when $f=0$.  In fact, Eqs.~(3) can be integrated explicitly in terms of elementary functions in each of the two regions $D_{On}$, $D_{Off}$ and the resulting solutions can be matched across the boundaries to obtain the full solution as given below. The solution is given by $ [x (t; t_0, x_0, y_0), ~y(t; t_0, x_0, y_0)]^T$ for which the initial condition is written as $ (t, x, y) $ $ = (t_0, x_0, y_0) $. Since Eq.~(3) is piecewise linear, the solution in each of the two regions can be obtained explicitly as follows.

\subsection{$D_{On}$ region $( y \ge 1)$}

In this region, $ f(y)= \alpha (y-1) - 2 \delta $  and hence Eqs.~(3a)~and~(3b) become
\begin{subequations}
\begin{eqnarray}
\dot x & = & y , \\
\dot y & = & - (\alpha (y-1) - 2 \delta) - x + f sin( \omega t), 
\end{eqnarray}
\end{subequations}
Differentiating Eq.~(8a) with respect to time and using Eqs.~(8a,~8b) 
in the resultant equation, we obtain,
\begin{eqnarray}
{\ddot x} + {A \dot x} + Bx = f sin( \omega t) +  (\alpha + 2 \delta)
\end{eqnarray}
where, $A = \alpha $ and $B = 1$. The roots of equation (9) is given by,
\begin{equation}
{m_{1,2}} =  \frac{-(A) \pm \sqrt{(A^{2}-4B)}} {2}, \\
\end{equation}
Since the roots $m_{1,2}$ of equation (9) are real and unequal, the general solution to Eq.(9) can be written as,  
\begin{equation}
x(t) = C_1 e^ {m_1 t} + C_2 e^ {m_2 t} + E_1 \sin \omega_1 t + E_2 \cos \omega_1 t + E_3
\end{equation}
where $C_1$, $C_2$ are the integration constants and
\begin{eqnarray}
E_1 & = & \frac {f (B-\omega^2)  }{A^2 \omega^2 + (B-\omega ^2)^2} \nonumber \\
E_2 & = & \frac {-f A \omega}{A^2 \omega^2 + (B-\omega^2)^2}\nonumber \\
E_3 & = & (\alpha + 2 \delta) \nonumber
\end{eqnarray}
Differentiating Eq.(11) and using it in Eq. (8a) we get, 
\begin{equation}
y(t) = {\dot{x}}
\end{equation} 
The constants $C_1$ and $C_2$ in the above equations can be evaluated by solving both Eqs.(11) and (12) for $C_1$ and $C_2$ at a suitable initial instant $t_0$, with $x_0$ and $y_0$ as initial conditions at time $t=t_0$, provided the trajectory of the dynamical system just enters the region $D_{On}$ at time $t_0$. The constants $C_1$ and $C_2$ can be obtained from Eqs.(11) and (12) as,
\begin{eqnarray}
C_1 =  \frac{e^ {- m_1 t_0}} {m_1 - m_2} ((y_0 - m_{2} x_0+ m_2 E_3) + (m_2 E_2 - \omega E_1) \cos \omega t_0) + (\omega E_2 + m_2 E_1) \sin \omega t_0) \nonumber \\
C_2 =  \frac{e^ {- m_2 t_0}} {m_2 - m_1} ((y_0 - m_{1} x_0+m_1 E_3) + (m_1 E_2 - \omega E_1) \cos \omega t_0) + (\omega E_2 + m_1 E_1) \sin \omega t_0) \nonumber
\end{eqnarray}

\subsection{$ D_{Off}$ region $( x < 1)$}

In the  $D_{Off}$ region the characteristic function is chosen as $f(y) = - 2 \delta y $. On substituting the above value of $f(y)$ in Eq.~(3a), we obtain 
\begin{subequations}
\begin{eqnarray}
\dot x & = & y , \\
\dot y & = & - 2 \delta y - x  + f sin( \omega t), 
\end{eqnarray}
\end{subequations}
Differentiating Eq. (13a) with respect to time and using Eqs. (13a, 13b) in the resultant equation, we obtain, 
\begin{eqnarray}
{\ddot x} + {C \dot x} + D x = f sin( \omega t)  
\end{eqnarray}
where $C =- 2 \delta$ and $D = 1$. The roots of Eq.(14) are a pair of complex conjugates given by $m_{3,4}$ = $ u \pm i v$ with
\begin{eqnarray*}
u & = & - \frac{C}{2}\\
v & = & \frac{ \sqrt{4D-C^2} }{2}
\end{eqnarray*}
The general solution to Eq.(14) is then given as, 
\begin{equation}
x(t) = e^{u t}(C_3 \cos(vt) + C_4 \sin(vt)) + E_4 \sin \omega_1 t + E_5 \cos \omega_1 t
\end{equation}
where $C_3$ and $C_4$ are the integration constants and 
\begin{eqnarray*}
E_4 & = & \frac {f (D-\omega^2)  }{C^2 \omega^2 + (D-\omega^2)^2}\\
E_5 & = & \frac {f C \omega}{C^2 \omega^2 + (D-\omega^2)^2}\\
\end{eqnarray*}
Differentiating Eq.(15) and using it in Eq. (13a) we get, 
\begin{equation}
y(t) = {\dot{x}}
\end{equation} 
The constants $C_3$ and $C_4$ in the above equations can be evaluated by solving both Eqs.(15) and (16) for $C_3$ and $C_4$ at a suitable initial instant $t_0$, with $x_0$ and $y_0$ as initial conditions at time $t=t_0$, provided the trajectory of the dynamical system just enters the region $D_{Off}$ at time $t_0$. The constants $C_3$ and $C_4$ can be obtained from Eqs.(15) and (16) as,
\begin{eqnarray}
C_3 =  \frac{e^{- u t_0}} {v} (( x_0 v \cos v t_0 + (u x_0 - y_0) \sin v t_0) + ((\omega E_4 - u E_5 )\sin vt_0 - v E_5 \cos v t_0 ) \sin \omega t_0 \nonumber \\ 
- ((\omega E_4 - u E_5 )\sin vt_0 + v E_4 \cos v t_0 ) \cos \omega t_0) \nonumber \\
C_4 =  \frac{e^{- u t_0}} {v} (( x_0 v \sin v t_0 + (y_0 - u x_0) \cos v t_0) - ((\omega E_4 - u E_5 )\cos vt_0 + v E_5 \sin v t_0 ) \cos \omega t_0 \nonumber \\ 
+ ((\omega E_5 + u E_4 )\cos vt_0 - v E_4 \sin v t_0 ) \sin \omega t_0) \nonumber
\end{eqnarray}

Now let us briefly explain how the solution can be generated in the $(x-y)$ phase space. Thus if we start with the initial conditions $x(t=0) = x_0, y(t=0) = y_0$ in the region $D_{On}$ region at time $t=0$, the arbitrary constants $C_1$ and $C_2$ get fixed. Then $x(t)$ evolves as given in Eq.(11), up to either $t=T_1$, when $x(T_1)=1$ and $\dot{x}(T_1) > 0$ or $t=T^{'}_1$, when $x{(T^{'}_1}) = -1$ and $\dot{x}(T^{'}_1) < 0$. Knowing whether $T_1 < T^{'}_1$ or $T_1 > T^{'}_1$ we can determine the next region of interest $(D_{Off})$ and the arbitrary constants of the solutions of that region can be fixed by matching the solutions. The procedure can be continued for each successive crossing. In this way, the explicit solutions can be obtained in each of the regions $D_{On}$, $D_{Off}$. However, it is clear that sensitive dependence on initial conditions is introduced in each of these crossings at appropriate parameter regimes during the inverse procedure of finding $T_1, T^{'}_1, T_2, T^{'}_2,...,$ etc. from the solutions.\\
\begin{figure}[H]
\begin{center}
\includegraphics[scale=0.55]{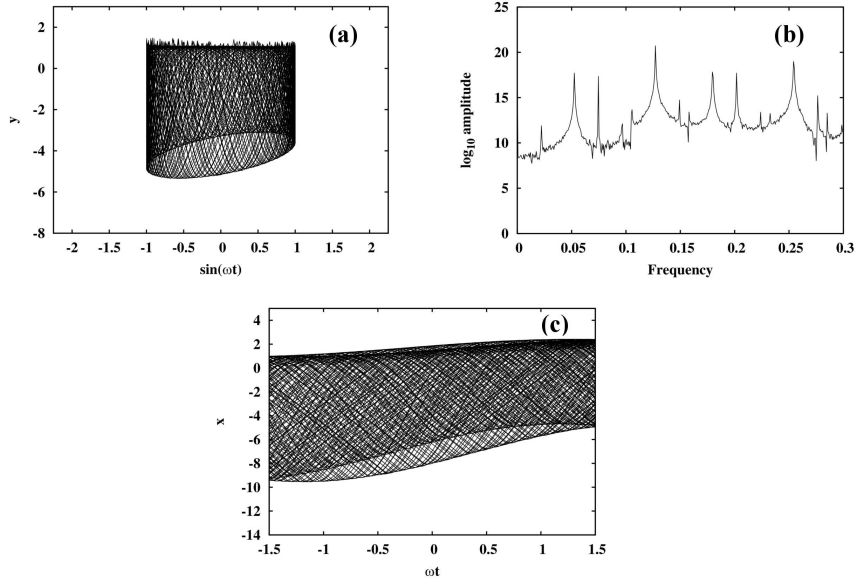}
\caption{Torus for $f=0.7318$ (a) phase portrait in $(sin (\omega t) - y)$ phase plane (b) power spectrum (c) phase portrait in $(\omega t-x)$ phase plane}
\label{fig2}
\end{center}
\end{figure}
\begin{figure}[H]
\begin{center}
\includegraphics[scale=0.55]{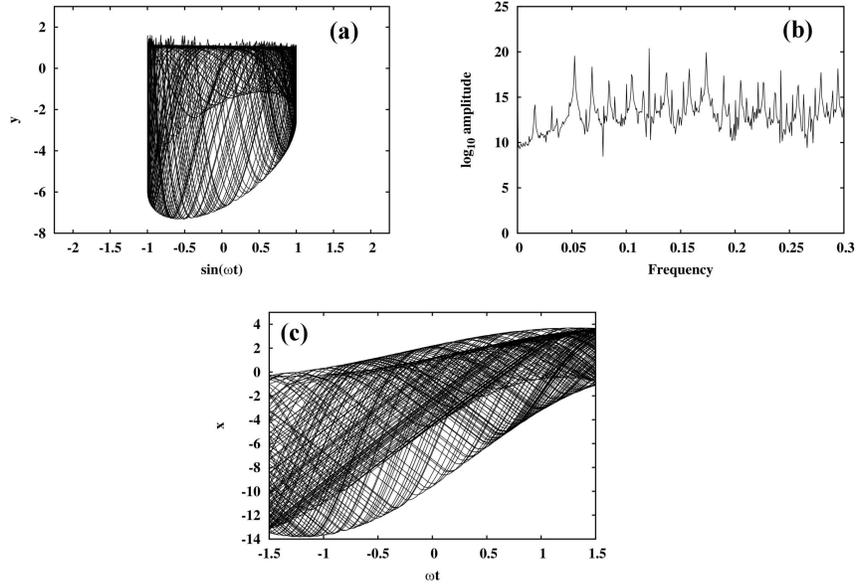}
\caption{Chaotic attractor for $f=2.017$ (a) phase portrait in $(sin (\omega t) - y)$ phase plane (b) power spectrum (c) phase portrait in $(\omega t-x)$ phase plane}
\label{fig3}
\end{center}
\end{figure}
\begin{figure}[H]
\begin{center}
\includegraphics[scale=0.55]{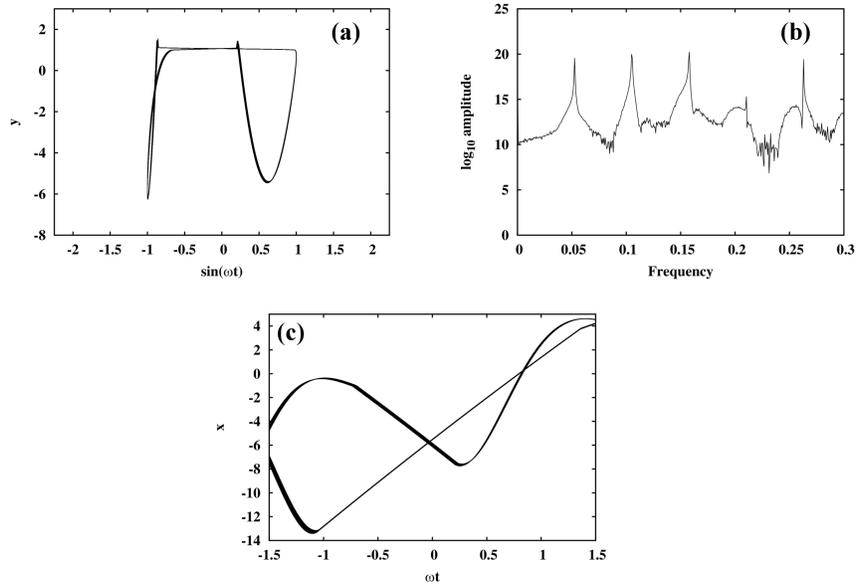}
\caption{Periodic solution for $f=2.95$ (a) phase portrait in $(sin (\omega t) - y)$ phase plane (b) power spectrum (c) phase portrait in $(\omega t-x)$ phase plane}
\label{fig4}
\end{center}
\end{figure}
Using the above analytical solutions the phase portraits of the circuit equations can be drawn to show torus breakdown or quasiperiodic route to chaos. The amplitute of the external periodic force is varied and the rescaled values of the circuit parameters are kept at $\delta=0.4, \omega=0.3$ and $\alpha=100$. A torus is observed for $f=0.7318$ which is as shown in Fig.\ref{fig2}. The phase portrait of the torus in $(sin(\omega t)-y)$ and $(\omega t-x)$ phase planes are shown in Fig.\ref{fig2}a, Fig.\ref{fig2}c and the power spectrum of the torus in Fig.\ref{fig2}b. On increasing the control parameter $f$, the torus breaks down to a chaotic attractor for $f=2.017$. The phase portraits and the power spectrum of the chaotic attractor are as shown in Fig.\ref{fig3}. Further increase in the control parameter leads to a limit cycle periodic solution as shown in Fig.\ref{fig4}.

\section{Conclusion}
In this paper, we presented the analytical results of a forced parallel $LCR$ circuit circuit with a diode and negative conductance. The torus breakdown or quasiperiodic route to chaos exhibited by the circuit is studied both numerically and experimentally but not studied analytically. An explicit analytical solution to the normalized circuit equations is presented and the quasiperiodic route to chaos is observed through phase portraits and power spectrum, obtained from the solution presented.

%

\end{document}